\documentclass[a4paper]{spie}  
\usepackage{amsmath,amsfonts,amssymb,aas_macros, url}
\usepackage{graphicx}
\usepackage[colorlinks=true, allcolors=blue]{hyperref}
\title{CUSP: a two cubesats constellation for Space Weather and solar flares X-ray polarimetry}
\author[a]{Sergio Fabiani}
\author[b]{Ilaria Baffo}
\author[c]{Sergio Bonomo}
\author[d]{Gessica Contini}
\author[a]{Enrico Costa}
\author[c]{Giovanni Cucinella}
\author[e]{Giovanni De Cesare}
\author[a]{Ettore Del Monte}
\author[d]{Andrea Del Re}
\author[a]{Sergio Di Cosimo}
\author[c]{Simone Di Filippo}
\author[a]{Alessandro Di Marco}
\author[b]{Pierluigi Fanelli}
\author[a]{Fabio La Monaca}
\author[f]{Alfredo Locarini}
\author[a]{Pasqualino Loffredo}
\author[a]{Giovanni Lombardi}
\author[a]{Gabriele Minervini}
\author[f]{Dario Modenini}
\author[a]{Fabio Muleri}
\author[c]{Andrea Negri}
\author[c]{Massimo Perelli}
\author[a]{John Rankin}
\author[a]{Alda Rubini}
\author[a]{Paolo Soffitta}
\author[f]{Eliseo Strollo}
\author[f]{Paolo Tortora}
\author[d]{Alessandro Zambardi}
\affil[a]{INAF-IAPS, via del Fosso del Cavaliere 100, 00133 Rome, Italy}
\affil[b]{DEIM, Universit\'a degli studi della Tuscia, Largo dell'Universit\'a, 01100 Viterbo, Italy}
\affil[c]{IMT s.r.l., via Carlo Bartolomeo Piazza 30, 00161 Rome, Italy}
\affil[d]{SCAI Connect s.r.l.,  Via Francesco Gentile 135, 00173 Rome, Italy}
\affil[e]{INAF-OAS Bologna, via Piero Gobetti 93/3, 40129 Bologna, Italy}
\affil[f]{Alma Mater Studiorum Universit\'a di Bologna - Department of Industrial Engineering and Interdepartmental Center for Industrial Aerospace Research, Via Fontanelle 40, 47121 Forl\'i, Italy}

\authorinfo{Further author information: (Send correspondence to Sergio Fabiani)\\Sergio Fabiani: E-mail: sergio.fabiani@inaf.it, Telephone: +39 06 4993 4450}
\begin{document} 
\maketitle
\begin{abstract}
The CUbesat Solar Polarimeter (CUSP) project aims to develop a constellation of two CubeSats orbiting the Earth to measure the linear polarisation of solar flares in the hard X-ray band by means of a Compton scattering polarimeter on board of each satellite. CUSP will allow to study the magnetic reconnection and particle acceleration in the flaring magnetic structures.
CUSP is a project approved for a Phase A study by the Italian Space Agency in the framework of the Alcor program aimed to develop CubeSat technologies and missions.
\end{abstract}
\section{INTRODUCTION}
\label{sec:intro}  
Solar flares (SFs) are violent energetic phenomena taking place on our Sun that can have a strong impact on the human activities both on ground and in space. Solar activity, including SFs, can degrade radio communications, can cause radio blackouts and interference with GPS and satellite communications. Moreover, high-energy particles (protons and electrons) can release their energy in the satellite electronics producing malfunctions and also the loss of the satellite. 
The occurrence of SFs is also linked to Coronal Mass Ejection (CME) and Solar Energetic Particle (SEPs)  events on the ground \cite{Papaioannou2016}. CUSP outcomes are intended to contribute to the present and future networks for Space Weather, including the future ASI SPace weather InfraStructure (ASPIS) \cite{Plainaki2018}. Polarimetry gives instantaneous clues about the flaring event. As soon as photons are detected, they can be downloaded and promptly analysed.
CUSP is currently approved for a phase A study by the Italian Space Agency in the framework of the Alcor program aimed to develop CubeSat technologies and missions. CUSP is one of the 20 selected missions among 49 proposals submitted to a call for CubeSat missions that involved 22 participants from Research Institutes and Universities and 78 companies, mainly Small and medium-sized enterprises (SMEs) 
Over the next years, the Agency plans to deploy them  in orbit.
INAF-IAPS is the Prime Contractor and the responsible for the scientific payload development which front-end and back-end electronics will be designed and realized by SCAI Connect s.r.l company. 
The platform, compatible with the CubeSat 6U standard, will be designed and produced by IMT s.r.l. The Interdepartmental Center for Aerospace Industrial Research (CIRI-AERO) of the University of Bologna is responsible for the mission analysis, while the University of Viterbo ``La Tuscia" will take care of the ground segment with the ground station located in the University Campus. 
\section{CUSP science objective and scientific requirements}
In the classical picture of a SF, magnetic reconnection originates the huge release of energy at the top of  a magnetic loop. Particles are accelerated along the magnetic field lines towards the lower layers of the solar atmosphere and the interplanetary space. The dominant components of the energy spectrum are:
\begin{itemize}
\item thermal Bremsstrahlung (expected weakly polarised) \cite{Emslie1980a} 
and emission lines below 10 keV
\item non-thermal Bremsstrahlung from about 10--20 keV, that is expected to be highly polarised \cite{Zharkova2010}
\end{itemize}
Theoretical models predict high linear polarisation in the X-rays, depending on the particle beaming and magnetic field properties \cite{Zharkova2010,Jeffrey2020}.
Moreover, the directivity of accelerated particles can be derived from polarisation measurements. Until nowadays, few measurements with low significance have been performed \cite{Tindo1970,Tindo1972a,Tindo1972b,Tramiel1984,SuarezGarcia2006,Boggs2006}.
High significance measurements would allow to overcome degeneracies in particle beaming models resulting from other observables like energy spectra \cite{Jeffrey2020}.
To perform a step forward in the understanding of the SF physics, CUSP will measure the linear polarisation in the 20-100 keV energy band with a Minimum Detectable polarisation $<10\%$ at least for SFs most relevant in terms of Space Weather (X class).
Moreover, due to the fact that SFs are dynamical events (with time scales from minutes to hours), CUSP will have the capability to study polarisation as a function of time. A detailed assessment of mission requirements will be carried out during phase A and B.
CUSP science program also inlcude to perform some ancillary science. CUSP can detect intense X-ray astrophysical sources falling in the Field of View during the year while observing the Sun (F.O.V. $\pm 21^\circ$ at the Sun):
\begin{itemize}
\item Gamma Ray Bursts: based on the SWIFT/BAT catalog, CUSP is expected to detect 16 GRBs/yrs with a peak flux over the absorber background
\item other sources as for example Crab Nebula (PWN), Sco X-1 (LMXB) and A0535+26 (HMXB)
\end{itemize}
\section{The payload: The hard X-ray polarimeter}
\begin{figure} [ht]
\begin{center}
\begin{tabular}{c} 
\includegraphics[height=5cm]{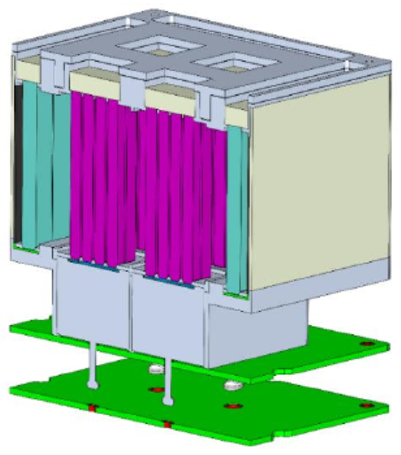}
\includegraphics[height=8cm]{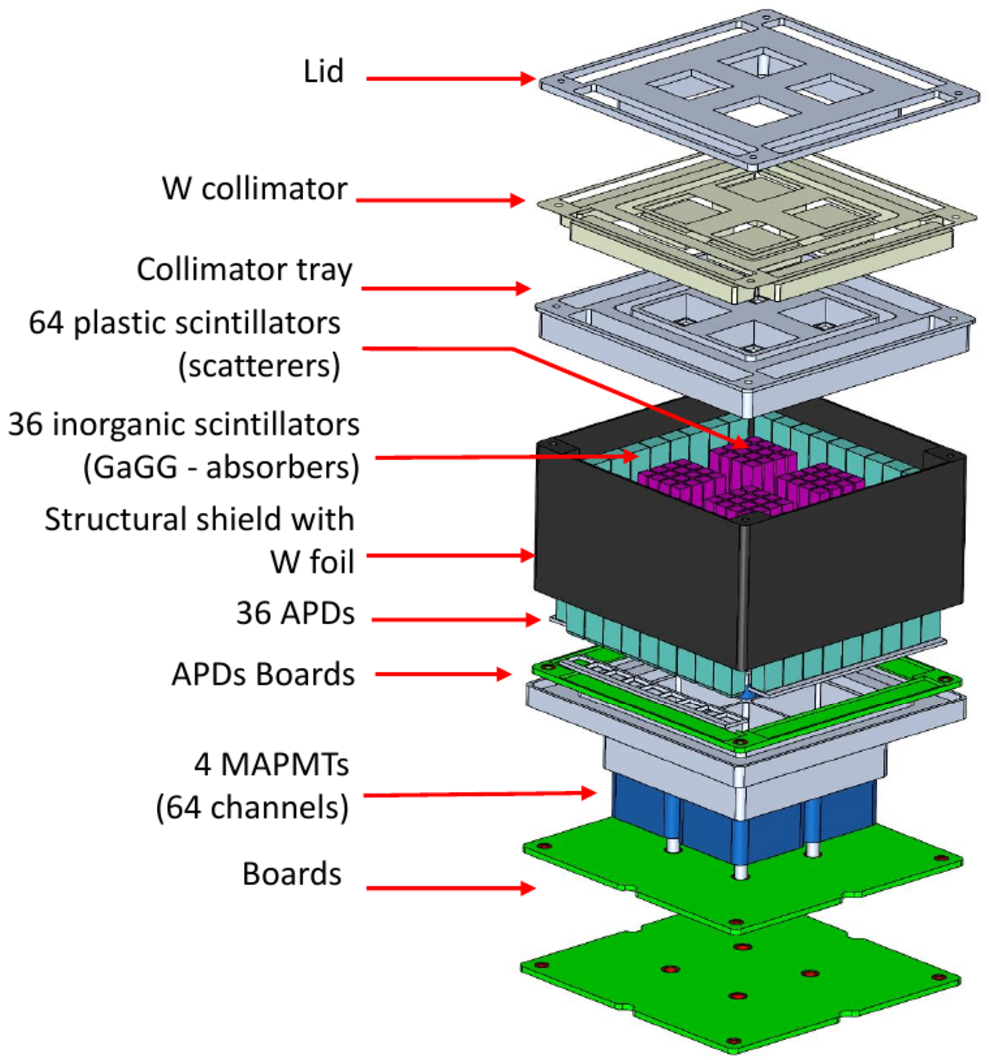}
\end{tabular}
\end{center}
\caption[example] 
{Compact (on the left) and exploded view of the hard X-ray polarimeter of the CUSP payload. \label{fig:views} }
\end{figure}

The payload hosts a dual-phase Compton scattering polarimeter (operating in the 20-100 keV energy band) that exploits coincidence measurements between plastic and inorganic scintillator rods. Fig.~\ref{fig:views} shows the compact and exploded view of the polarimeter. The low atomic number of the plastic scintillator allows to maximise the scattering probability with respect to the heavier inorganic crystal made of GaGG (Gd3Al2Ga3O12) that maximises the photoelectric absorption of the scattered photon. Polarimetry is performed by measuring the azimuthal angular distribution of plastic/GaGG. If radiation is polarised, the preferential angular direction of polarisation produces a preferential response of the detector in the azimuthal direction (normal to the incident beam axis) as described by the Klain-Nishina cross section for Compton scattering \cite{Heitler1954}: 
\begin{equation}
\frac{d\sigma}{d\Omega}=\frac{{r_0}^2}{2}\frac{{E^\prime}^2}{{E}^2}\Biggr[ \frac{E}{E^\prime}+\frac{E^\prime}{E}-2\sin^2 \theta \cos^2 \phi \Biggl] \label{eq:KN}
\end{equation}
where 
 \begin{equation}
\frac{E'}{E}=\frac{1}{1+\frac{E}{m_e c^2}(1-\cos \theta)}\label{eq:EsuE}
\end{equation}
The energies of the incident and scattered photons are $E$ and $E^\prime$, respectively. The angle $\theta$ is the scattering angle measured from the incident photon direction and $\phi$ is the azimuthal angle measured from the plane containing both the incoming direction and the electric vector of the incident photon. Linearly polarised photons are preferentially scattered perpendicularly to their polarisation direction.
Thus, the response of $\phi$ emission directions for a polarised beam is modulated. 
The higher the modulated response, the higher the sensitivity of the polarimeter.
Therefore, the modulation factor $\mu(\theta)$ (fraction of modulated signal corresponding to 100$\%$ polarised radiation) is:
\begin{equation}
\mu(\theta)=\frac{N_\mathrm{max}(\theta)-N_\mathrm{min}(\theta)}{N_\mathrm{max}(\theta)+N_\mathrm{min}(\theta)}=\frac{(\frac{d\sigma}{d\Omega})_{\phi=\frac{\pi}{2}}-(\frac{d\sigma}{d\Omega})_{\phi =0}}{(\frac{d\sigma}{d\Omega})_{\phi=\frac{\pi}{2}}+(\frac{d\sigma}{d\Omega})_{\phi =0}}=\frac{\sin^2\theta }{\frac{E}{E^\prime}+\frac{E^\prime}{E}-\sin^2 \theta} \label{eq:Muphi}
\end{equation} 
In the limit of coherent scattering  $E = E^\prime$, the maximum modulation factor is for $\theta=90^\circ$ (orthogonally to the incident photons direction). By increasing the energy it occurs at narrower scattering angles (forward folding effect). However, at 100 keV it is still $\theta \simeq 90^\circ$ \cite{Fabiani2012c}.

\begin{figure} [ht]
\begin{center}
\begin{tabular}{c} 
\includegraphics[height=5cm]{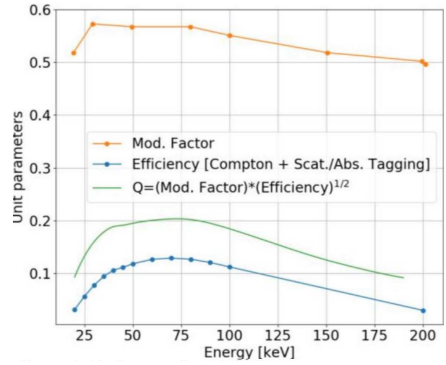}
\end{tabular}
\end{center}
\caption[example] 
{Current best estimates of the modulation factor, efficiency (Compton interaction and tagging efficiency) and quality factor. \label{fig:curves}}
\end{figure}

The payload of CUSP comprises a W collimator for limiting the field of view about $\pm 21^\circ$ around the solar direction, the plastic and the inorganic scintillators assemblies which are readout by means of 4 Multi-anodes Photomultiplier Tubes (MAPMTs) for a total of 64 channels and 36 Avalanche Photo-diodes (APDs), respectively. Readout sensors have been selected to have a high heritage due to the very short implementation time required for the project.
They are the MAPMT R7600 (rugged version to survive launch vibration) and an SMD version of APD derived from the S8664-55, both by Hamamatsu. 
MAPMTs and APDs are readout by MAROC 3A and SKIROC 2A ASICs by WEEROC.
The payload also comprises A/D conversion, Micro HVs and a payload computer to handle House Keepings.
In Fig.~\ref{fig:curves} are shown the current best estimates of the modulation factor $\mu$ (response in terms of modulation for 100$\%$ polarised radiation), efficiency $\epsilon$ (Compton interaction and tagging efficiency) and quality factor $Q$. Tagging efficiency is the probability to detect an event in the scatterer after a detection of an event in the absorber \cite{Fabiani2012c}. The quality factor is:
 \begin{equation}
Q=\mu \sqrt{\epsilon}\label{eq:Q}
\end{equation}
It identifies the energy range in which the polarimeter is effective in measuring polarisation.
This parameter is derived from the Minimum Detectable Popularization (MDP)\cite{Weisskopf2010} by assuming a source dominated observation (if background is negligible). This is true especially for brighter SFs.
Table\ref{tab:mdp} reports the current best estimate of the MDP based on benchmark solar flares from Saint-Hilaire et al. (2008)\cite{SaintHilaire2008}.
Few minutes of integration time allow to measure the polarisation of solar flares that is expected to be well above the MDP (at a level of some tens of per cent).
\begin{table}[ht]
\caption{Current best estimate of the MDP based on benchmark solar flares from Saint-Hilaire et al. (2008)\cite{SaintHilaire2008}} 
\label{tab:mdp}
\begin{center}       
\begin{tabular}{|c|c|c|} 
\hline
\rule[-1ex]{0pt}{3.5ex} Flare Class & Integration Time (s) & MDP ($\%$)  \\
\hline
\hline
\rule[-1ex]{0pt}{3.5ex} M 5.2 & 284 & 9.2  \\
\hline
\rule[-1ex]{0pt}{3.5ex} X 1.2 & 240 & 4.8  \\
\hline
\rule[-1ex]{0pt}{3.5ex} X 10 & 351 & 0.9  \\
\hline
\end{tabular}
\end{center}
\end{table} 
\section{The mission concept}
\begin{figure} [ht]
\begin{center}
\begin{tabular}{c} 
\includegraphics[height=5cm]{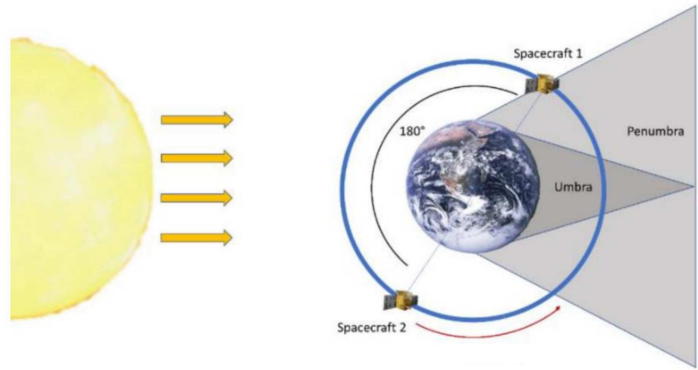}
\end{tabular}
\end{center}
\caption[example] 
{Constellation of two CubeSats at 180$^\circ$ phase difference along the orbit. \label{fig:orbit}}
\end{figure}

\begin{figure} [ht]
\begin{center}
\begin{tabular}{c} 
\includegraphics[height=5cm]{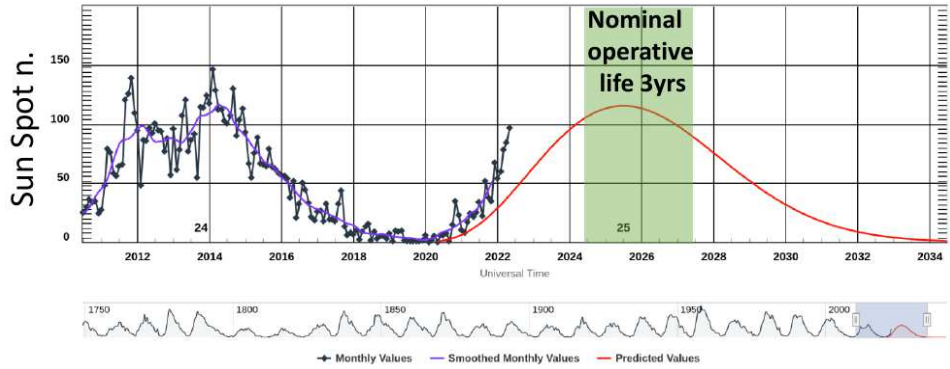}
\end{tabular}
\end{center}
\caption[example] 
{Prediction of the 25th solar cycle activity in terms of sunspot number in relation with the 3 years nominal operative life of the CUSP mission.  \label{fig:solarcycle}}
\end{figure}

We foresee a constellation consisting of 1 orbital plane and 2 CubeSats at 180$^\circ$ of phase difference along the orbit (see Fig.~\ref{fig:orbit}). This configuration allows to have always at least 1 satellite of the constellation in daylight for any $\beta$ angle (orientation of the orbital plane with respect to the Sun) for observing the Sun. Depending on the specific orbit, there is also a partial overlap of different duration between the two observation phases of the two satellites. 
A preliminary mission analysis identifies as a target a sun-synchronous (SSO) orbit at an altitude between 500 and 600 km with local time ascending node LTAN variable between a Mid-Morning scenario (LTAN = 9:30) and a Noon-Midnight scenario (LTAN = 12:00). This range of orbits offers the largest number of launch opportunities and foresees different orientations of the orbital plane with respect to the Sun with different duration of the eclipse period.
Both satellites will be at the same inclination and altitude to have a common orbital period and node precession. 
The maximum contemporary daylight for both satellites is about 16 minutes per orbit for an SSO Mid-Morning (LTAN = 9:30) at 600 km of altitude.
The time scale of the project, until launch, and then the operational phase of the mission, are compatible with the observation of the Sun during the next cycle maximum. 
Fig.~\ref{fig:solarcycle} reports the prediction of the 25th solar cycle activity by NOAA\footnote{https://www.swpc.noaa.gov/products/solar-cycle-progression} in terms of sunspot number and over-imposed the 3 years nominal operative life of the CUSP mission.
The peak of the solar activity and, therefore, the maximum occurrence of solar flares, is expected between 2024 and 2028.
The CUSP mission foresees two main operative modes. The observation mode requires that the each satellite
rotates around the direction of pointing towards the Sun, a direction aligned with the axis of symmetry of the sensitive elements of the polarimeter. The rotation reduces the systematic effects (spurious modulation) induced by the polarimeter geometry. 
The duration of peak intensity in the hard X-rays of a solar flare is of the order of a few minutes, thus the rotation speed of 1RPM is sufficient for sampling the modulation curve induced by polarisation on a time scale of 30 seconds (half of the rotation period).
The rotation of the polarimeter is obtained with the rotation of the entire satellite. The payload foresees a data storage and pre-processing system on board to archive data before their transmission to the ground. During the downlink phase, the satellite will change its attitude, optimizing the transmission link, thus interrupting the observation phase. The downlink of the data will take place by ensuring the continuity of observation of the solar phenomena of interest, in the context of the constellation of two satellites.
\section{The platform}
\begin{figure} [ht]
\begin{center}
\begin{tabular}{c} 
\includegraphics[height=5cm]{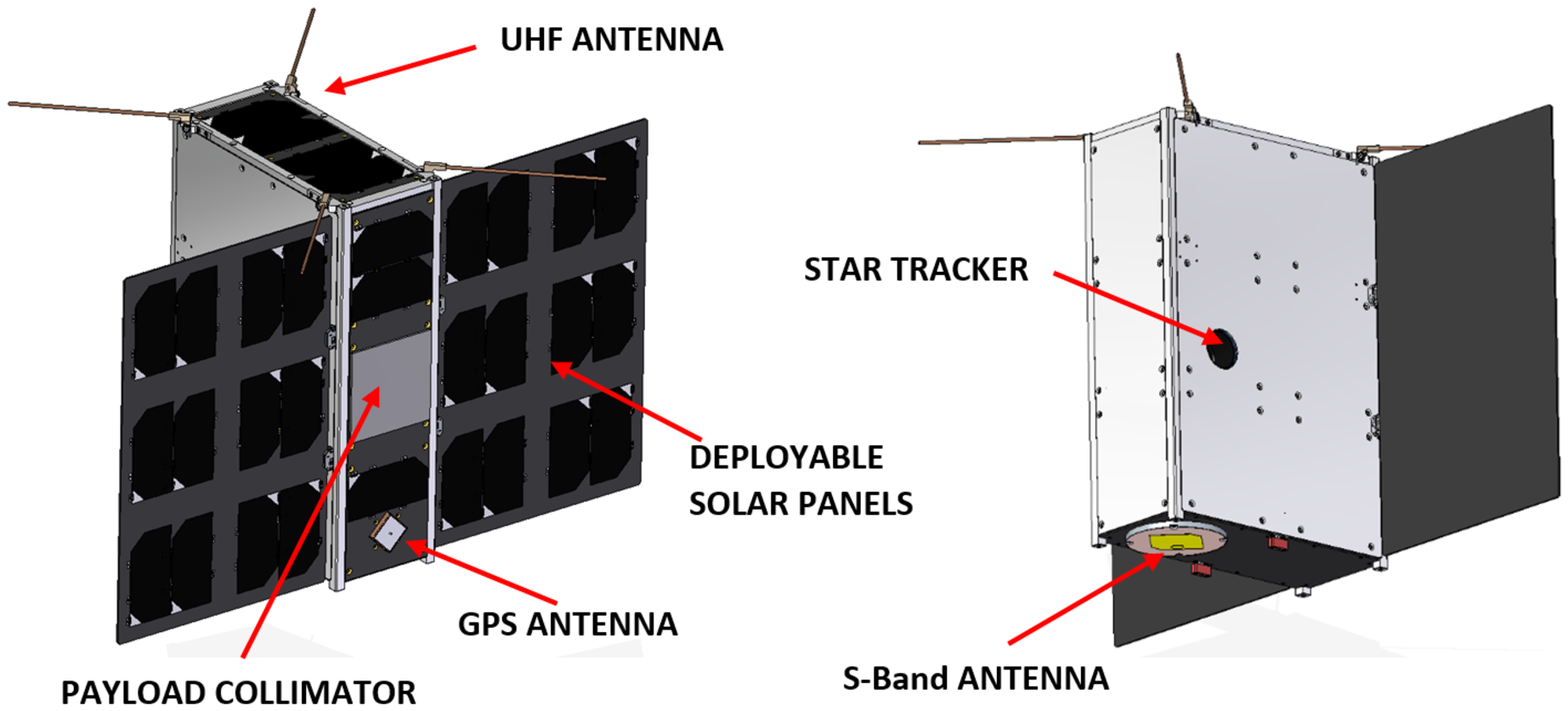} \\
\includegraphics[height=5cm]{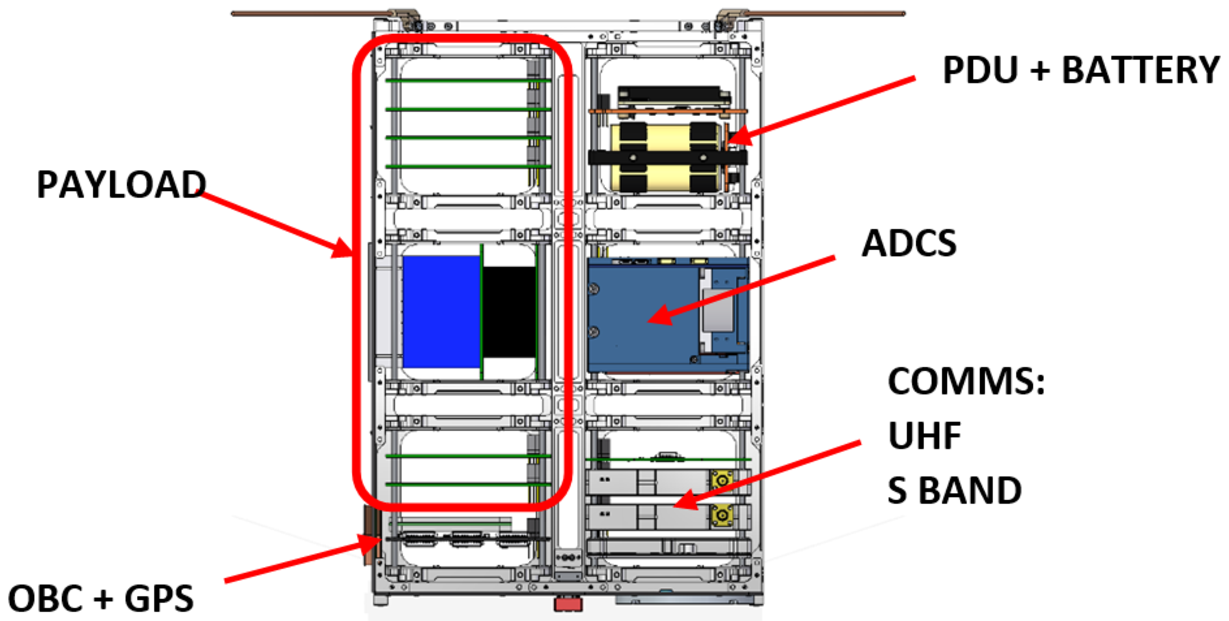}
\end{tabular}
\end{center}
\caption{CUSP platform. The top images show the external view. The bottom image shows the allocation of subsystems including the payload. \label{fig:platform}}
\end{figure}

\begin{figure} [ht]
\begin{center}
\begin{tabular}{c} 
\includegraphics[height=11cm]{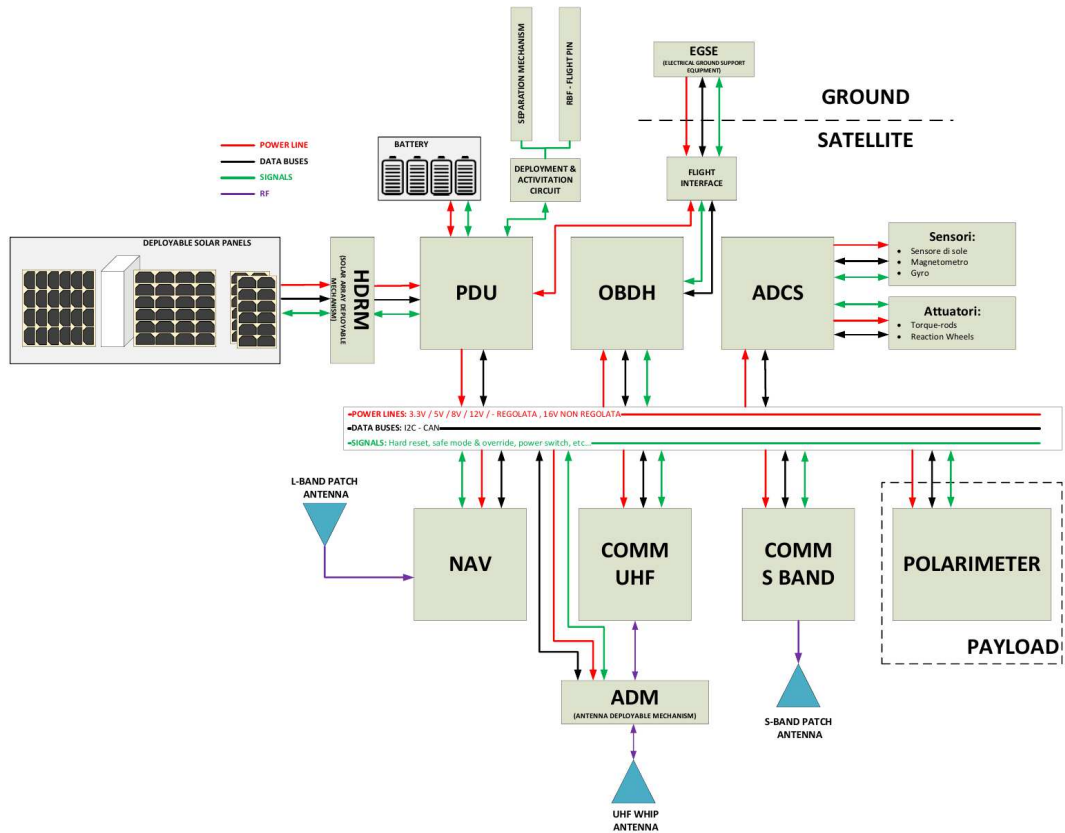} \\
\end{tabular}
\end{center}
\caption{Scheme of subsystems of a CUSP satellite. \label{fig:platformsub}}
\end{figure}

The platform foreseen for the CUSP project (see Fig.~\ref{fig:platform}) is a standard 6U CubeSat architecture developed by IMT s.r.l., that has a consolidated experience in the realization of different CubeSats, also in the institutional framework. The 6U CubeSat is based on the heritage of the HORTA and EOSS platforms (6U CubeSat platforms funded by Italian regional POR / FESR 2014-20 projects of Lazio and Puglia regions, respectively).
The preliminary performance of the CUSP platform are reported in Table~\ref{tab:platform}
\begin{table}[ht]
\caption{Preliminary performance of the CUSP platform.} 
\label{tab:platform}
\begin{center}       
\begin{tabular}{|c|c|} 
\hline
\rule[-1ex]{0pt}{3.5ex} Peak Power & 30W with deployable panels  \\
\hline
\hline
\rule[-1ex]{0pt}{3.5ex} Battery & $>$75Wh   \\
\hline
\rule[-1ex]{0pt}{3.5ex} Pointing accuracy & $>$ 1$^\circ$ (down to 0.1$^\circ$ )   \\
\hline
\rule[-1ex]{0pt}{3.5ex} Operative frequencies & S / UHF   \\
\hline
\rule[-1ex]{0pt}{3.5ex} Downlink  throughput & Up to 5 Mbps   \\
\hline
\rule[-1ex]{0pt}{3.5ex} Available interfaces & CAN Bus, I2C, UART, RS422   \\
\hline
\rule[-1ex]{0pt}{3.5ex} Regulated bus & 3,3V, 5V e 12V   \\
\hline
\rule[-1ex]{0pt}{3.5ex} Not regulated bus & 32V (24V-33.6V) or 16V (12V-16.8V)   \\
\hline
\rule[-1ex]{0pt}{3.5ex} Available volume for the payload & 2.5U   \\
\hline
\rule[-1ex]{0pt}{3.5ex} Nominal life time & 3 years in LEO   \\
\hline
\end{tabular}
\end{center}
\end{table} 
The platform hosts an On Board Computer that underwent a qualification for radiation hardness above 20 krad and implements mitigation strategies for latch-up errors (SEU / SEL) by architecture, adopting anti-latchup and TMR (Triple Modular Redundancy) circuit. It can also offer memory storage for some Payload data (thanks to NAND memories) with the possibility to manage independently the Payload.
During the 1RPM rotation required by the payload, the satellite will provide the determination of the attitude with an accuracy better than 0.5$^\circ$, needed for processing the scientific data. This will be achieved thanks to a high-precision gyroscopes and star trackers.
During the non-observational phases of the Payload, the platform returns to the three axes stabilized mode to perform the downlink of the collected data towards the ground station.
The telemetry and remote controls are realized through UHF band uplink and downlink communication with
omnidirectional turnstile antennas. UHF antenna deployment will be based on the consolidated design by IMT s.r.l. which has already implemented several deployment mechanisms, both of antennas and solar panels. 
The S-band communication subsystem, in support of the UHF one, provides a higher link
speed to allow scientific data transfer.
The power needed by the on-board equipment is guaranteed by body mounted and deployable solar panels up to about 30W. The PDU subsystem will provide for the regulation and distribution of on-board energy.
Fig.~\ref{fig:platformsub} shows the scheme of a CUSP satellite subsystems.
\section{The ground station}
\begin{figure} [ht]
\begin{center}
\begin{tabular}{c} 
\includegraphics[height=5cm]{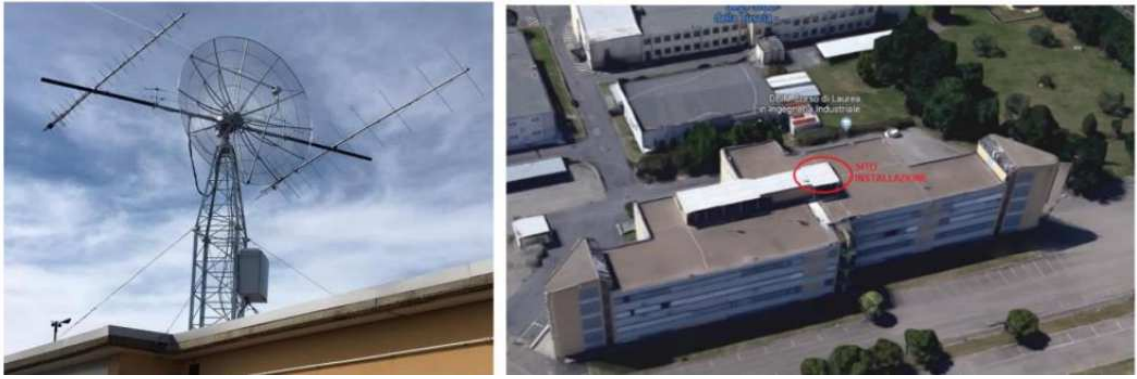}
\end{tabular}
\end{center}
\caption{The ground station at the “La Tuscia” University of Viterbo. \label{fig:groundstation}}
\end{figure}

\begin{figure} [ht]
\begin{center}
\begin{tabular}{c} 
\includegraphics[height=5cm]{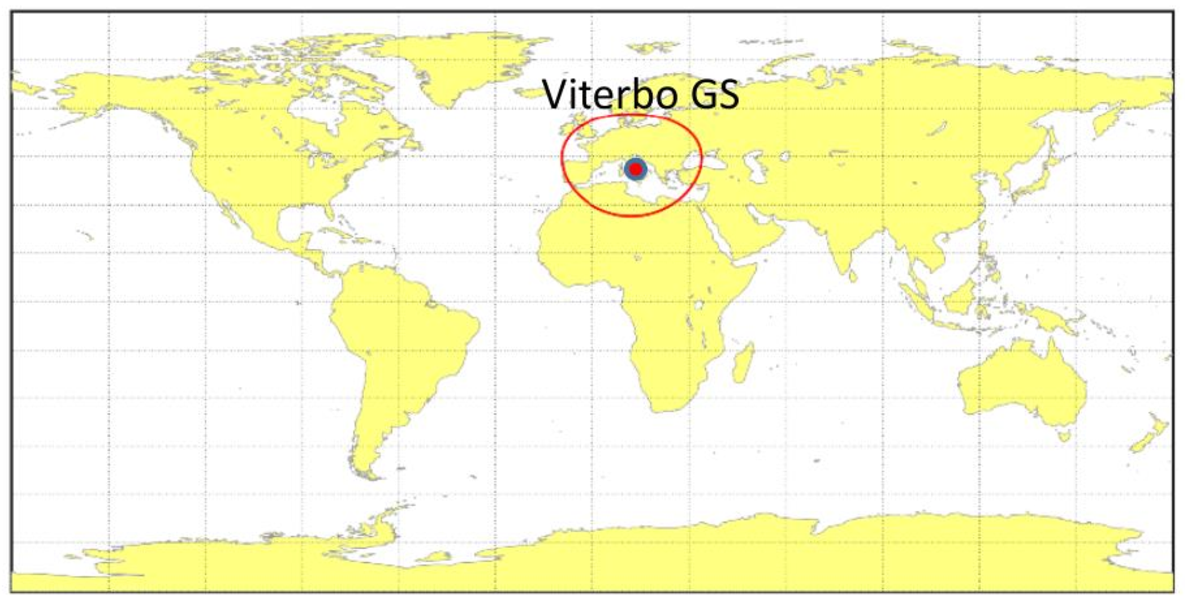}
\includegraphics[height=5cm]{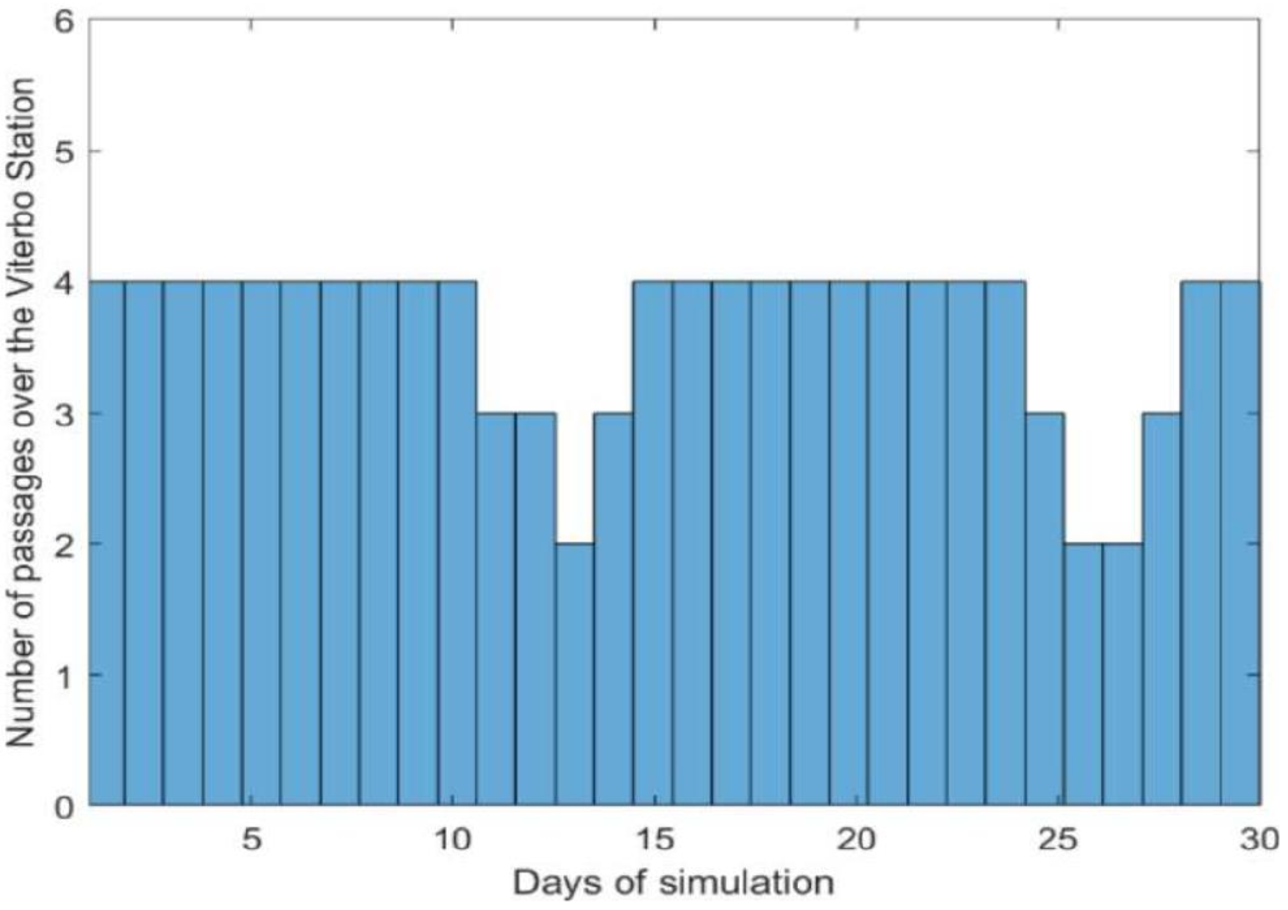}
\end{tabular}
\end{center}
\caption{Left panel: location of the ground station on the Earth. Right panel: preliminary estimation of the number of contacts with the ground station.\label{fig:contacts}}
\end{figure}

The Ground Station (see Fig.~\ref{fig:groundstation}) is located at the “La Tuscia” University of Viterbo, on the building F of the ``Riello" Campus at the following coordinates (see left panel of Fig.~\ref{fig:contacts}):
\begin{itemize}
\item Latitude = 42.413$^\circ$ N
\item Longitude = 12.113$^\circ$ E
\item Minimum elevation $\epsilon$ min = 10$^\circ$
\end{itemize}
It was built in 2019 as part of the HORTA project (Italian regional funds POR-FESR 2014-2020 of Lazio region). The Station allows for autonomous satellite tracking (using TLE satellite data - Two Lines Elements) and  satellite communication. In addition, it can be controlled remotely.
Available antennas and bands are:
\begin{itemize}
\item VHF: Uplink and Downlink
\item UHF: Uplink and Downlink
\item S-band: Downlink
\end{itemize}
The UHF/VHF bandwidth are 9.6 kbps as default for downlink (available also 1.2/ 2.4 / 4.8 kbps)
and 1.2 kbps as default for uplink (available also 2.4 / 4.8/ 9.6 kbps).
The S-band bandwidth is up to 1 Mbps for downlink.
The pointing accuracy of the ground station is 0.1° (both azimuth and elevation) with a minimum tracking speed of 2$^\circ$/sec in azimuth, 1.8$^\circ$/sec in elevation.
Data received from the ground station is transferred via fibre optics cable to dedicated workstations in the Mission Control Center that allows to schedule the passage of the satellite, the
TT\&C and Payload data transceiving activities. Moreover, it provides a Network Server service for data delivery to third parties. 
From the preliminary Mission Analysis, about 2-4 contacts per day will be possible during mission operations (see right panel of Fig.~\ref{fig:contacts})
\section{The planning}
The CUSP project is based on subsystems with high TRL. The Platform and the Ground Station can exploit a large heritage that allow them to guarantee a TRL 7. The Payload  is based on elements with a high TRL (MAPMT, APD, scintillators, ASICs, coincidence technique), but the polarimeter as whole needs to be implemented. Thus, a TRL 3 is quoted.
The Model Philosophy is based on the production of 1 detector prototype at the end of Phase B. It will be representative of the detector front-end to enhance detector TRL from 3 to 4.
Then 1 payload EQM will be designed during phase B to be produced and tested during phase C. it will be representative of the payload to enhance detector TRL from 4 to 7.
From the satellite point of view, 2 CubeSats will be produced:
\begin{itemize}
\item 1 Proto-flight Model (PFM). To be qualified at proto-qualification level
\item 1 Flight Model (FM). To be qualified at acceptance level.
\end{itemize}
The Calibration of the Hard X-ray Polarimeter of each CubeSat will be carried out at INAF-IAPS calibration facility (already employed for calibrating the IXPE Detector Units)\cite{Muleri2021ice}.
A 3 months phase A (starting in September 2022) plus possibly a 12 months phase B were approved.
\section{Conclusions}
Our Sun is still an astrophysical source with many aspects not well understood. Solar Flare mechanisms of production and particle acceleration are still debated. Moreover, solar flares can represent a threat for human technological activities in space and on ground, because they are usually correlated to Solar Energetic Particles Events (SEPs) at the Earth and Coronal Mass Ejections (CMEs). 
CUSP mission is aimed to measure the linear polarisation of solar flares in the 20-100 keV energy band to probe of particle acceleration and magnetic field behaviour during such energetic events.
X-ray polarimetry allows to measure particle directivity and magnetic filed structure of the flaring loop to assess also magnetic reconnection that is thought to be at the origin of solar flares.
CUSP will contribute to the understanding of these solar violent phenomena also participating in the present and future networks for Space Weather, including the ASI SPace weather InfraStructure (ASPIS).

\begin{thebibliography}{10}

\bibitem{Papaioannou2016}
{Papaioannou}, A., {Sandberg}, I., {Anastasiadis}, A., {Kouloumvakos}, A.,
  {Georgoulis}, M.~K., {Tziotziou}, K., {Tsiropoula}, G., {Jiggens}, P., and
  {Hilgers}, A., ``{Solar flares, coronal mass ejections and solar energetic
  particle event characteristics},'' {\em Journal of Space Weather and Space
  Climate}~{\bf 6},  A42 (Dec. 2016).

\bibitem{Plainaki2018}
{Plainaki}, C., {Negri}, B., {Castronuovo}, M., and {Antonelli}, A., ``{Towards
  an Italian Space Weather Infrastructure: the ASPIS project},'' in [{\em EGU
  General Assembly Conference Abstracts}{\nolinebreak\hspace{0.1em}]},  {\em
  EGU General Assembly Conference Abstracts},  5239 (Apr. 2018).

\bibitem{Emslie1980a}
{Emslie}, A.~G. and {Brown}, J.~C., ``{The polarization and directivity of
  solar-flare hard X-ray bremsstrahlung from a thermal source},'' {\em
  \apj}~{\bf 237},  1015--1023 (May 1980).

\bibitem{Zharkova2010}
{Zharkova}, V.~V., {Kuznetsov}, A.~A., and {Siversky}, T.~V., ``{Diagnostics of
  energetic electrons with anisotropic distributions in solar flares. I. Hard
  X-rays bremsstrahlung emission},'' {\em \aap}~{\bf 512},  A8 (Mar. 2010).

\bibitem{Jeffrey2020}
{Jeffrey}, N. L.~S., {Saint-Hilaire}, P., and {Kontar}, E.~P., ``{Probing solar
  flare accelerated electron distributions with prospective X-ray polarimetry
  missions},'' {\em \aap}~{\bf 642},  A79 (Oct. 2020).

\bibitem{Tindo1970}
{Tindo}, I.~P., {Ivanov}, V.~D., {Mandel'Stam}, S.~L., and {Shuryghin}, A.~I.,
  ``{On the Polarization of the Emission of X-Ray Solar Flares},'' {\em
  \solphys}~{\bf 14},  204--207 (Sept. 1970).

\bibitem{Tindo1972a}
{Tindo}, I.~P., {Ivanov}, V.~D., {Mandel'Stam}, S.~L., and {Shuryghin}, A.~I.,
  ``{New Measurements of the Polarization of X-Ray Solar Flares},'' {\em
  \solphys}~{\bf 24},  429--433 (June 1972).

\bibitem{Tindo1972b}
{Tindo}, I.~P., {Ivanov}, V.~D., {Valn{\'{\i}}{\v c}ek}, B., and {Livshits},
  M.~A., ``{Preliminary Interpretation of the Polarization Measurements
  Performed on 'Intercosmos-4' during Three X-Ray Solar Flares},'' {\em
  \solphys}~{\bf 27},  426--435 (Dec. 1972).

\bibitem{Tramiel1984}
{Tramiel}, L.~J., {Novick}, R., and {Chanan}, G.~A., ``{Polarization evidence
  for the isotropy of electrons responsible for the production of 5-20 keV
  X-rays in solar flares},'' {\em \apj}~{\bf 280},  440--447 (May 1984).

\bibitem{SuarezGarcia2006}
{Suarez-Garcia}, E., {Hajdas}, W., {Wigger}, C., {Arzner}, K., {G{\"u}del}, M.,
  {Zehnder}, A., and {Grigis}, P., ``{X-Ray Polarization of Solar Flares
  Measured with Rhessi},'' {\em \solphys}~{\bf 239},  149--172 (Dec. 2006).

\bibitem{Boggs2006}
{Boggs}, S.~E., {Coburn}, W., and {Kalemci}, E., ``{Gamma-Ray Polarimetry of
  Two X-Class Solar Flares},'' {\em \apj}~{\bf 638},  1129--1139 (Feb. 2006).

\bibitem{Heitler1954}
{Heitler}, W.,  [{\em {Quantum theory of
  radiation}}{\nolinebreak\hspace{0.1em}]}, International Series of Monographs
  on Physics, Oxford: Clarendon, 1954, 3rd ed. (1954).

\bibitem{Fabiani2012c}
{Fabiani}, S., {Campana}, R., {Costa}, E., {Del Monte}, E., {Muleri}, F.,
  {Rubini}, A., and {Soffitta}, P., ``{Characterization of scatterers for an
  active focal plane Compton polarimeter},'' {\em Astroparticle Physics}~{\bf
  44},  91--101 (Apr. 2013).

\bibitem{Weisskopf2010}
{Weisskopf}, M., {Elsner}, R., and {O'Dell}, S., ``{On understanding the
  figures of merit for detection and measurementof x-ray polarization},'' {\em
  Proceedings of the SPIE,}~{\bf 7732},  77320E--77320E--5 (2010).

\bibitem{SaintHilaire2008}
{Saint-Hilaire}, P., {Krucker}, S., and {Lin}, R.~P., ``{A Statistical Survey
  of Hard X-ray Spectral Characteristics of Solar Flares with Two
  Footpoints},'' {\em \solphys}~{\bf 250},  53--73 (July 2008).

\bibitem{Muleri2021ice}
{Muleri}, L. and {et al.}, ``{The IXPE Instrument Calibration Equipment},''
  {\em In Preparation}~{\bf 00},  00--00 (Dec. 2021).

\end{thebibliography}
\bibliographystyle{spiebib} 

\end{document}